\documentclass[%
nofootinbib,
 amsmath,amssymb,
 aps,
 prd,
floatfix,notitlepage,
twocolumn, superscriptaddress,
]{revtex4-1}
\usepackage[normalem]{ulem}
\usepackage{slashed}
\usepackage{orcidlink}
\usepackage{graphicx}
\usepackage{dcolumn}
\usepackage{bm}
\usepackage{hyperref}
\usepackage{footnote}
\usepackage{hyperref}
\usepackage{xcolor}
\begin{document}

\title{Power corrections to the photon polarization tensor in a hot and dense medium of massive fermions}

\author{Osvaldo {\sc Ferreira}\orcidlink{0000-0002-6711-8308}}
 \affiliation{
 Instituto de F\'\i sica, Universidade Federal do Rio de Janeiro,\\
 CEP 21941-972 Rio de Janeiro, RJ, Brazil 
}

\author{Eduardo S. {\sc Fraga}\orcidlink{0000-0001-5340-156X}}
\affiliation{
 Instituto de F\'\i sica, Universidade Federal do Rio de Janeiro,\\
 CEP 21941-972 Rio de Janeiro, RJ, Brazil 
}

\date{\today}
\begin{abstract}
 We compute the $\mathcal{O}(k^2)$ terms (power corrections) of the photon polarization tensor in a hot and dense medium of particles with a small but finite mass, i.e., $0< m\ll T, \mu$. We perform our calculations within the hard thermal loop approximation in the real-time formalism, and evaluate the first nonzero mass corrections. For a renormalization scale $\bar{\Lambda}\sim T$, these mass contributions determine the temperature dependence of the power corrections. These results have direct implications in the computation of electric and magnetic susceptibilities of hot and dense media in equilibrium. We address such implications and make comparisons with previous results.  
\end{abstract}

\maketitle


\section{Introduction}

The photon polarization tensor (or self-energy) is a basic building block for numerous calculations in high-energy physics \cite{Peskin:1995ev}. Its evaluation in a medium at finite temperature ($T$) and chemical potential ($\mu$) is a necessary step in the description of hot plasmas and dense matter \cite{Weldon:1982,Altherr:1992mf, Blaizot:1995kg,Manuel:1995td}. In particular, it is relevant in the study of collective phenomena, particle production, and to the computation of transport coefficients \cite{Thoma:2008my, Thoma:1995ju, Carrington:1997sq, Schmitt:2017efp}. 

Efforts to evaluate the one-loop photon polarization tensor at finite temperature within the hard thermal loop (HTL) approximation \cite{Laine:2016book, Kapusta, Bellac:2011kqa, Blaizot:2001nr, Ghiglieri:2020} can be traced back to Ref. \cite{Toimela:1984xy}. Those results were extended to two loops for high-temperature plasmas \cite{Carignano:2017}, and recently generalized to finite chemical potentials \cite{Gorda:2022, Ekstedt:2023oqb}. The leading order (LO) HTL evaluation of the photon self-energy corresponds to the first-order contribution of an expansion in terms of a soft external momentum. The inclusion of corrections coming from higher powers of the soft external momentum represents a way to go beyond the LO result (complementary to computing more loops), and can be relevant to the evaluation of thermodynamic properties of a dense QED medium \cite{Gorda:2022zyc}. 

In this paper we compute the $\mathcal{O}(k^2)$ terms (power corrections) of the photon polarization tensor in a hot and dense medium of particles with a small but finite mass, i.e., $0< m\ll T, \mu$. We perform our calculations within the hard thermal loop approximation in the real-time formalism, and evaluate the first nonzero mass corrections. As we will show, for a renormalization scale $\bar{\Lambda}\sim T$, these mass contributions determine the temperature dependence of the power corrections.

In the high-density regime, the case of massive fermions has been investigated in Ref. \cite{Altherr:1992jg}, where the authors found analytic expressions for the one-loop photon self-energy and spectral density (at LO) for a degenerate QED system ($T=0$), using the imaginary-time formalism. In Ref. \cite{Stetina:2017ozh}  these results were rederived within the real-time formalism, and applied to the study of the photon propagation in dense nuclear matter with possible implications to neutron star physics. 

Recently, effects from small fermion masses on the photon self-energy at finite temperature and density were considered within the real-time formalism in Ref.  \cite{Comadran:2021}. The authors evaluated mass corrections to the leading-order photon polarization tensor and showed that these corrections can be as important as power and loop corrections, depending on the fermion mass and the external momentum compared to $e T$ (with $e$ being the gauge coupling). Such mass corrections can be relevant for the energy loss of fermions propagating through a hot medium \cite{Comadran:2023vsr}. Reference \cite{Haque:2018eph} also investigated relevant thermodynamic quantities for QCD with massive quarks within the imaginary-time formalism.

For soft momenta, power corrections can give contribution as important as two-loop corrections \cite{Carignano:2021zhu, Carignano:2017, Carignano:2019ofj, Manuel:2016wqs}, which makes them crucial for a precise evaluation of thermodynamic quantities of hot and dense QED \cite{Gorda:2022, Gorda:2022zyc}. Moreover, these corrections have a direct impact on the electric and magnetic susceptibilities of hot plasmas in equilibrium \cite{Weldon:1982, Endrodi:2022}. Nevertheless, this connection could only be indirectly inferred by previous work.

The electric susceptibility of a hot plasma has been recently revisited in Refs. \cite{Endrodi:2022, Endrodi:2023wwf}, inspired by previous work by Weldon \cite{Weldon:1982}. By requiring that the system should be in equilibrium, the authors of Refs. \cite{Endrodi:2022, Endrodi:2023wwf} defined the susceptibility in a way that is treatable within lattice simulations. Previous efforts \cite{Elmfors:1994fw,Elmfors:1998ee,Loewe:1991mn, Gies:1998vt, Gies:1999xn} based on the use of the Schwinger propagator \cite{Schwinger:1951nm} lead to different results. The disagreement is speculated to be due to a different ordering of limits when performing the weak-field expansion (see the discussion in Ref. \cite{Endrodi:2022}). Here, we investigate the inclusion of nonzero fermion masses on the computation of power corrections and its impact on electromagnetic susceptibilities, adopting a procedure that is similar to the one introduced in Ref. \cite{Comadran:2021}.

This paper is organized as follows. In Section \ref{section2} we introduce conventions that will be used throughout the paper, as well as quantities that will be the building blocks of our calculation. In Section \ref{section3} we present the main calculation, the evaluation of the first nonzero mass corrections to the components of the photon polarization tensor. In Section \ref{section4} we discuss its implications for the electric and magnetic susceptibilities of hot and dense systems, and compare with previous results and lattice simulations. In Section \ref{section5} we summarize our findings and make some final remarks.

\section{Photon polarization tensor within the real-time formalism}\label{section2}

\subsection{Notation}

We adopt, with minor adaptations, the notation used in Ref. \cite{Gorda:2022}. Our calculations are performed in $D=4-2\epsilon$ dimensions and $d=3-2\epsilon$ spatial dimensions. In the real-time formalism we have the Minkowski metric, and we choose the signature $g_{\mu \nu}=\mathrm{diag}(-1, +1, +1, +1)$. Four-vectors are denoted as $P\equiv (p_{0}, \mathbf{p})$. Also $p=|\mathbf{p}|$ and $p^{i}$ ($i=1, 2,...,d$) denote individual spatial components. Our units are such that $c=\hbar=k_{B}=1$. 
We perform the calculations for the photon polarization tensor using the Keldysh basis representation for the real-time formalism \cite{Keldysh:1964, Chou:1984}. The reader is referred to Refs. \cite{Ghiglieri:2020, Gorda:2022} for more details. The labels attributed to some specific integrals used here are similar but not identical to those used in Ref. \cite{Gorda:2022}. They are listed in Appendix \ref{integrals}, which also contains more details on the notation and the use of dimensional regularization.
\subsection{Photon polarization tensor}
\begin{figure}
    \centering
    \includegraphics[height=3cm]{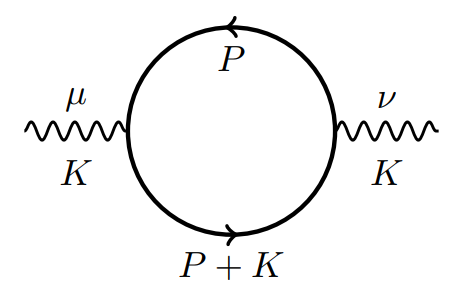}
    \caption{Feynman diagram for the one-loop photon  polarization tensor.}
    \label{vacuum_polarization}
\end{figure}

In this section we make some general statements about the polarization tensor and its power corrections, setting the ground for the calculations that follow. Within the real-time formalism the one-loop retarded polarization tensor (see Fig. 1) is given by \cite{Gorda:2022}
%
\begin{equation}\label{1_loop_Photon_polarization_tensor}
   \begin{aligned}
        -i\left(\Pi^R\right)_{\mu \nu}(K)&=-\int_P F_{\mu \nu}\left[ N_F^{-}(P)+N_F^{+}(P)\right]\\
    &\times\Delta^R(K+P) \Delta^d(P)\,, 
   \end{aligned}   
\end{equation}
where $P$ is the loop momentum and $K$ is the external momentum of the photon. Here, $N_F^{ \pm}(P)=\frac{1}{2}-n_F\left(p^0 \pm \mu\right)$, where $n_{F}$ is the Fermi-Dirac distribution function. The $F_{\mu \nu}$ factor is given by
\begin{equation}
\begin{aligned}
    F_{\mu \nu}(K, P)&=-4 e^2[2 P_\mu P_\nu+K_\mu P_\nu+P_\mu K_\nu \\
    &-\left(P^2+K \cdot P+m^2\right) g_{\mu \nu}]\,,
\end{aligned}
\end{equation}
where we assume a free electron-photon vertex function, $e$ being the coupling constant. The function
 \begin{equation}
     \Delta^{R/A}=\frac{-i}{P^2+m^2\pm i\eta p^{0}}
 \end{equation}
corresponds to the scalar part of the retarded ($R$) and advanced ($A$) fermion propagator, $S^{R/A}(P)=-(\slashed{P}+m)\Delta^{R/A}(P)$. The constant $\eta$ is a small parameter and $m$ is the fermion mass. $\Delta^{d}$ is the spectral function, defined as
\begin{equation}
    \Delta^{d}=\Delta^R-\Delta^A=2 \pi sgn(p^{0})\delta (P^2+m^2)\,.
\end{equation}

The polarization tensor can be decomposed into longitudinal and transverse components
\begin{equation}\label{polarization_tensor_decomposition}
    \Pi^{\mu \nu}(K)=\mathcal{P}_{\mathrm{T}}^{\mu \nu}(K) \Pi_{\mathrm{T}}(K)+\mathcal{P}_{\mathrm{L}}^{\mu \nu}(K) \Pi_{\mathrm{L}}(K)\,,    
\end{equation}
where $\mathcal{P}_{\mathrm{T}}^{\mu \nu}(K)$ and $\mathcal{P}_{\mathrm{L}}^{\mu \nu}(K)$ are orthogonal projectors for the longitudinal and transverse components \cite{Gorda:2022, Laine:2016book, Kapusta}, and the scalar functions 
\begin{equation}\label{polarization_tensor_decomposition2}
    \Pi_{\text{L}}=\frac{-K^2}{k^2}\Pi_{00} \quad , \quad  \Pi_{\mathrm{T}}=\frac{1}{d-1}\left(\Pi^{\mu}_\mu + \frac{K^2}{k^2}\Pi_{00} \right)
\end{equation}
are the transverse and longitudinal components of the polarization tensor, respectively.

From Equation \eqref{polarization_tensor_decomposition} we see that the polarization tensor is given once $\Pi_{\mathrm{T}}$ and $\Pi_{\mathrm{L}}$ are determined. To evaluate these components, we must calculate $\Pi^{R}_{00}(K)$ and $ (\Pi^{R})^{\mu}_{\mu}(K)$. For the $00$-component of the polarization tensor, Equation \eqref{1_loop_Photon_polarization_tensor} gives
\begin{equation}\label{Pol_Tensor_00}
    \begin{aligned}
    \Pi^{R}_{00}(K)&=4 e^2 \int_P \Delta^d(P)\left[N_F^{-}(P)+N_F^{+}(P)\right]\\
    &\times\frac{2 p_0^2+2 k^0 p^0+K \cdot P}{2 K \cdot P+K^2}\,,
    \end{aligned}
\end{equation}
where we have taken the fermion field to be on-shell\footnote{As discussed in \cite{Gorda:2022}, $\Delta^d$ will assure that the particles are on-shell, and it is safe to assume it even after integrating over $p_{0}$.}, $P^2=-m^2$. We also have absorbed the $i\eta$ factor into $k_{0}$ to simplify the notation and render the calculations cleaner. At the end of the calculation we have to impose the retarded prescription, $k_{0}\rightarrow k_{0}+i\eta$, on our results. 

From now on we implicitly assume that we are performing the calculations for the retarded polarization tensor.
The contributions of order $k^2$ (the ``power corrections") are found by expanding the integrand of Equation \eqref{Pol_Tensor_00} for $K$ small\footnote{Although we have not checked this explicitly, we believe that once the $p_{0}$ integrals are done, the order in which the expansions in small $K$ and in small mass are performed does not change the final result. However, doing the small $K$ expansion first appears to be the more straightforward approach.}. Performing this expansion and collecting the $\mathcal{O}(k^2)$ terms yields the following expression:
\begin{widetext}
\begin{equation}\label{Pol_Tensor_Power_corrections}
\Pi^\mathrm{Pow}_{00}(K)=\frac{e^2}{2} \int_P \Delta^d(P)\left[N_F^{-}(P)+N_F^{+}(P)\right]
\left(\frac{K^4} {(K \cdot P)^2}+\frac{2 K^4 k^0 p^0}{(K \cdot P)^3}-\frac{K^6 p_0^2}{(K \cdot P)^4}\right).
\end{equation}

For the transverse component of the photon polarization tensor, in addition to $\Pi_{00}$ we need to compute the trace $\Pi^{\mu}_\mu$. In this work we are then interested in the power corrections to
\begin{equation}
    (\Pi^{R})^{\mu}_{\mu}=-4e^2\int_P \Delta^d(P)\left[N_F^{-}(P)+N_F^{+}(P)\right]\left[ \frac{(D-2)K\cdot P+2m^2}{K^2+K\cdot P}\right]\,,
\end{equation}
which are given by
\begin{equation}\label{power_corrections_trace}
    (\Pi^{\mathrm{Pow}})^{\mu}_{\mu}=\frac{e^2}{2}(D-2)\int_P \Delta^d(P)\left[N_F^{-}(P)+N_F^{+}(P)\right]\left[ -\frac{K^4}{(K\cdot P)^2}+\frac{m^2K^6}{2(K\cdot P)^4}\right].
\end{equation}

More specifically, we want to investigate how mass corrections affect $\Pi^\mathrm{Pow}_{00}(K)$ and $(\Pi^{\mathrm{Pow}})^{\mu}_{\mu}$. Notice that, from the start, we already have an additional term proportional to $m^2$ in $(\Pi^{\mathrm{Pow}})^{\mu}_{\mu}$ that is not present in previous calculations.

For the purpose of later reference, we state here the expressions for the power corrections in the massless case \cite{Gorda:2022}:
\begin{equation}\label{Transverse_massless}
 \Pi_{\mathrm{T}}^{\text {P0 }}=-\frac{e^2}{4 \pi^2} \frac{2 K^2}{3}\left\{\ln \frac{2 \mathrm{e}^{-\gamma_{\mathrm{E}}} T}{\bar{\Lambda}}-1+\frac{1}{4}+\left(1-\frac{K^2}{4 k^2}\right)\left[1-\frac{k^0}{2 k} \ln \frac{k^0+k+i \eta}{k^0-k+i \eta}\right]-\mathrm{Li}_0^{(1)}\left(-\mathrm{e}^{\frac{\mu}{T}}\right)-\mathrm{Li}_0^{(1)}\left(-\mathrm{e}^{-\frac{\mu}{T}}\right)\right\},
\end{equation}
%
\begin{equation}\label{Longituginal_massless}
 \Pi_{\mathrm{L}}^{\text {P0}}=-\frac{e^2}{4 \pi^2} \frac{2 K^2}{3}\left\{\ln \frac{2 \mathrm{e}^{-\gamma_{\mathrm{E}}} T}{\bar{\Lambda}}-1+\left(1+\frac{K^2}{2 k^2}\right)\left[1-\frac{k^0}{2 k} \ln \frac{k^0+k+i \eta}{k^0-k+i \eta}\right]-\mathrm{Li}_0^{(1)}\left(-\mathrm{e}^{\frac{\mu}{T}}\right)-\mathrm{Li}_0^{(1)}\left(-\mathrm{e}^{-\frac{\mu}{T}}\right)\right\}.
\end{equation}
where we have introduced the notation $\Pi_{\mathrm{L}}^{\text{P}n}$ where P stands for ``power correction'' and $n$ indicates that the contribution is proportional to $m^{2n}$. Some additional comments regarding this expressions are given in the next section.

\end{widetext}
 
\section{Mass corrections to $\mathcal{O}(k^2)$}\label{section3}
In this section we evaluate the contributions brought about by a nonzero, small mass ($m \ll T,\mu$ or, equivalently, $m \ll p$) to power corrections to the polarization tensor. We follow a procedure similar to the one introduced in Ref. \cite{Comadran:2021}. We also use dimensional regularization at intermediate steps, adopting the same conventions of Ref. \cite{Gorda:2022}. This allows us to make direct comparisons to their results. In the following calculations, the basic idea is to expand the functions of the loop momentum in powers of $\frac{m^2}{p^n}$ ($n > 1$) and integrate the resulting expressions. More concretely, this amounts to substituting the expansions
\begin{widetext}
\begin{align}\label{expansions_on_mass}
    \frac{1}{E^n_{p}}&=\frac{1}{p^{n}}-\frac{n}{2}\frac{m^2}{p^{n+2}}+\frac{n(n+2)}{8}\frac{m^4}{p^{n+4}}+\cdots\\
    \overline{N}_{F}(E_{p})&=\overline{N}_{F}(p)+\frac{m^2}{2p}\frac{d \overline{N}_{F}(p)}{dp} -\frac{m^4}{8 p^3}\frac{d \overline{N}_{F}(p)}{dp}+\frac{m^4}{8 p^2}\frac{d^2 \overline{N}_{F}(p)}{dp^2}+\cdots\\  
    \frac{1}{(v' \cdot K)}&=\frac{1}{(v \cdot K)^n}+\frac{n}{2}\frac{m^2}{p^2}\frac{kz}{(v \cdot K)^{n+1}}-\frac{3 n}{8}\frac{m^4}{p^4}\frac{kz}{(v \cdot K)^{n+1}}+\frac{n(n+1)}{8}\frac{m^4}{p^4}\frac{(kz)^2}{(v \cdot K)^{n+2}}+\cdots
\end{align}
\end{widetext}
in the expressions for $\Pi_{00}$ and $\Pi_{\mu}^{\mu}$, leading to integrals of the form
\begin{equation}
m^{2n}\int_{p}\frac{\overline{N}_{F}(p)}{p^{3+2n}}F(v\cdot K)\sim \frac{m^{2n}}{T^{2n}}f(k, k_{0})+\mathcal{O}(\epsilon) \,,
\end{equation}
where $n=1,2,...$. Here, $F(v\cdot K)$ is a function of the external momentum $K=(k, k_{0})$ and of the angular parameter $\mathbf{\hat{k}}\cdot\mathbf{\hat{p}}=z$ and $f(k, k_{0})$ is the result of the angular integrations.

\subsection{$\mathcal{O}(m^2)$ contribution}

We now evaluate terms proportional to $m^2$ that will lead, in the appropriate limits, to $\mathcal{O}(\frac{m^2}{T^2})$ and $\mathcal{O}(\frac{m^2}{\mu^2})$. We start by performing the integral over $p^0$ in Equation \eqref{Pol_Tensor_Power_corrections}, which gives
\begin{equation}\label{Power_corrections_nonzero_mass}
\begin{aligned}
    \Pi^{\text {Pow }}_{00}(K)&=\frac{e^2}{2} \int_{\mathbf{p}} \frac{1}{E_{p}^3}\overline{N}_{F}(E_{p})\bigg[\frac{K^4}{(v' \cdot K)^2}+\frac{2 k^0 K^4}{(v' \cdot K)^3}\\
        &-\frac{K^6}{(v' \cdot K)^4}\bigg]\,,
\end{aligned}
\end{equation}
where $v ' \cdot K=-E_{p} k_{0}+pkz$, with $z=\hat{p}\cdot\hat{k}$, and we have defined $\overline{N}_{F}\equiv N_F^{-}(P)+N_F^{+}(P)$. Notice that, since the denominators of the terms in the brackets are dependent on $E_{p}$, they should also be expanded. Substituting the expansions defined above in \eqref{Power_corrections_nonzero_mass} and keeping the $m^2$ terms, we obtain
\begin{equation}
\begin{aligned}
    \Pi^{\text {Pow}}_{00}&=\Pi^{\text {P0}}_{00}+\frac{e^2m^2}{4}\int_{\mathbf{p}}\frac{\overline{N}_{F}}{p^5}\left[f_{2}(v\cdot K)-3 f_{1}(v\cdot K)\right]\\
    &+\frac{e^2m^2}{4}\int_{\mathbf{p}}\frac{1}{p^4}\frac{d \overline{N}_{F}}{d p}f_{1}(v\cdot K)+\mathcal{O}(m^4)\,,
\end{aligned}
\end{equation}
where we have defined 
\begin{equation}
    f_{1}(v\cdot K)=\frac{K^4}{(v \cdot K)^2}+\frac{2 k^0 K^4}{(v \cdot K)^3}-\frac{K^6}{(v \cdot K)^4}
\end{equation}
and 
\begin{equation}
    f_{2}(v\cdot K)=2K^4\frac{kz}{(v \cdot K)^3}+6 k^0 K^4\frac{kz}{(v \cdot K)^4}-4K^6\frac{kz}{(v \cdot K)^5}\,.
\end{equation}

The term $\Pi^{\text {P0}}_{00}$ corresponds to the power corrections in the massless case. It has been computed previously using different methods \cite{Carignano:2017, Carignano:2019ofj, Carignano:2021zhu, Gorda:2022}. In the infrared regime (small $p$), it is finite due to a cancellation between matter and vacuum parts. In the ultraviolet regime (large $p$), its divergence is removed by a wave function renormalization. These facts have been shown in all references mentioned above. Due to a similarity in the conventions and notations, our results can be directly compared to the ones in Ref. \cite{Gorda:2022}. As will be shown below, the relevant integrals in this work are all finite. Nevertheless, we use dimensional regularization through the whole calculation to keep the mass corrections fully consistent with the massless terms. 
Using the relations from Appendix \ref{integrals} and denoting the $\mathcal{O}(m^2)$ terms $\Pi^{\text {P1}}_{00}$, we find that
\begin{equation}
    \begin{aligned}
    \Pi^{\text {P1}}_{00}&=\frac{e^2m^2}{4}\int_{p}\frac{\overline{N}_{F}}{p^5}\int_{z}\big[f_{2}(v\cdot K)-f_{1}(v\cdot K)\\
    &+2\epsilon f_{1}(v\cdot K)\big].
    \end{aligned}
\end{equation} 
Then, one can show that the angular integral over the difference $f_{2}(v\cdot K)-f_{1}(v\cdot K)$ is proportional to $\epsilon$ and, therefore,
\begin{equation}
    \Pi^{\text {P1}}_{00}(K)\overset{d\rightarrow 3}{=}0.
\end{equation}

From Equation \eqref{polarization_tensor_decomposition2}, we see that the power corrections do not give contributions proportional to $m^2$ to the longitudinal component of the polarization tensor. Notice that this result is valid for both $T$ and $\mu$ finite, and for nonzero $k$ and $k_{0}$.

Let us now evaluate the trace $(\Pi^{\mathrm{Pow}})^{\mu}_{\mu}$. Performing the integral over $p^0$ in Equation \eqref{power_corrections_trace}, we obtain
\begin{equation}
    \begin{aligned}
        (\Pi^{\mathrm{Pow}})^{\mu}_{\mu}&=\frac{e^2}{2}(d-1)\int_{\mathbf{p}} \frac{1}{E_{p}^3}\overline{N}_{F}(E_{p})\bigg[ -\frac{K^4}{(v'\cdot K)^2}\\
        &+\frac{m^2K^6}{2E_{p}^2(v'\cdot K)^4}\bigg]\,.
    \end{aligned}
\end{equation}
\vspace{0.24cm}

Using the expansions in Equation \eqref{expansions_on_mass}, one can show that the $\mathcal{O}(m^2)$ contribution is given by
\begin{equation}
   \begin{aligned}
     (\Pi^{\text{P}1})^{\mu}_{\mu}&=\frac{e^2m^2}{4}(d-1)\int_{\mathbf{p}} \frac{\overline{N}_{F}}{p^5}\left(\frac{K^2}{(v\cdot K)^4}+\frac{2 kz}{(v\cdot K)^3}-\frac{(d-2)}{(v\cdot K)^2} \right)\\
        &=\frac{e^2m^2}{4}(d-1)\mathcal{N}\mathcal{R}_{5}K^4\left(K^2\mathcal{A}_{4}-2k_{0}\mathcal{A}_{3}-\mathcal{A}_{2}-2\epsilon\mathcal{A}_{2}\right)\,,
   \end{aligned}
\end{equation}
where $\mathcal{N}$ is a normalization constant. Its definition and the expression for the integral $R_{5}$ are given in Appendix \ref{integrals}. Substituting the expressions for the integrals found in Appendix \ref{integrals}, we find, for $d\rightarrow 3$,
\begin{equation}
    (\Pi^{\text{P}1})^{\mu}_{\mu}=\frac{k^2e^2}{12 \pi^2}\frac{m^2}{T^2}\left(\text{Li}_{-2}^{(1)}(-e^{-\frac{\mu}{T}})+\text{Li}_{-2}^{(1)}(-e^{\frac{\mu}{T}})\right)\,,
\end{equation}
where we have used the notation $\text{Li}_{s}^{(1)}$ for the derivatives with respect to the order $s$ of the polylogarithm function and defined $\displaystyle\text{Li}_{s_{0}}^{(1)}(z)\equiv\lim_{s\rightarrow s_{0}} \frac{\partial\text{Li}_{s}(z)}{\partial s}$.
\\

\subsection{$\mathcal{O}(m^4)$ contribution}

 Since, as discussed in the previous section, for the $\mu=\nu=0$ component the contributions proportional to $m^2$ vanish, in this section we compute the $m^4$ contributions to the power corrections to the polarization tensor. These give a nonzero contribution, as shown below. 
 
 Let us consider again Equation \eqref{Power_corrections_nonzero_mass}.
Using the expansions \eqref{expansions_on_mass}, one can show that the $\mathcal{O}(m^4)$ contributions are of the form
\begin{widetext}
    \begin{equation}
    \begin{aligned}
    \Pi^{\text {P2}}_{00}&=\frac{e^2m^4}{8}\int_{p}\frac{\overline{N}_{F}}{p^7}\int_{z}\left\{ \left[\frac{1}{2}(d+1)(d-7)+15/2\right]f_{1}(v\cdot K)-(d-4)f_{2}(v\cdot K)+f_{3}(v\cdot K)\right\}\\
    &=\frac{e^2m^4}{16}\int_{p}\frac{\overline{N}_{F}}{p^7}\int_{z}\left[2f_{3}(v\cdot K)-f_{1}(v\cdot K)+ \mathcal{O}(\epsilon)\right]\,,
    \end{aligned}
\end{equation}
where
\begin{equation}
    f_{3}(v\cdot K)=-3 K^4\frac{kz}{(v \cdot K)^3}+3K^4\frac{(kz)^2}{(v \cdot K)^4}-9 k_{0} K^4\frac{kz}{(v \cdot K)^4}\\
    +12 k_{0}K^4\frac{(kz)^2}{(v \cdot K)^5}+6K^6\frac{kz}{(v \cdot K)^5}-10 K^6 \frac{(kz)^2}{(v \cdot K)^6}\,.
\end{equation}
Performing the angular integrals, we arrive at (after a fair amount of algebra) 
\begin{equation}
        \Pi^{\text {P2}}_{00}(K)=\frac{e^2 m^4}{4}\mathcal{N}\mathcal{R}_{7}k^2\left(1+8\frac{k^2k_{0}^2}{(K^2)^2}\right)\,.
\end{equation}
Substituting the radial integral and taking $d\rightarrow 3$, we find 
\begin{equation}
        \Pi^{\text {P2}}_{00}(K)\overset{d\rightarrow 3}{=}\frac{e^2m^4k^2}{16\pi^2}\left(\frac{\text{Li}^{(1)}_{-4}(-e^{\frac{-\mu}{T}})+\text{Li}^{(1)}_{-4}(-e^{\frac{\mu}{T}})}{24 T^4}\right)\left(1+8\frac{k^2k_{0}^2}{(K^2)^2}\right).
\end{equation}

For completeness, we also show our results for the $\mathcal{O}(m^4)$ contributions to the trace. Considering expression \eqref{power_corrections_trace} again, and performing the appropriate expansions we arrive at
\begin{equation}
   \begin{aligned}
     (\Pi^{\text{P2}})^{\mu}_{\mu}&=-\frac{e^2m^4}{2}(d-1)\mathcal{N}\mathcal{R}_{7} K^4\left[ -\frac{1}{8}\mathcal{A}_{2}-\frac{1}{4}\left(\mathcal{A}_2+k_{0}\mathcal{A}_{3}+\frac{3}{4}\mathcal{A}^{(2)}_{4}\right)\right]\\
        &+\frac{e^2m^4}{4}\mathcal{N}\mathcal{R}_{7} K^6\left(\frac{3}{2}\mathcal{A}_{4}+2k_{0}\mathcal{A}_{5}\right)\,.
   \end{aligned}
\end{equation}
Substituting the integrals from Appendix \ref{integrals}, we find
\begin{equation}
    (\Pi^{\text{P2}})^{\mu}_{\mu}=-\frac{K^2e^2}{128\pi^2}\frac{m^4}{T^4}\left(\text{Li}^{(1)}_{-4}(-e^{\frac{-\mu}{T}})+\text{Li}^{(1)}_{-4}(-e^{\frac{\mu}{T}})\right)\left(1-\frac{(5k^4+k_{0}^4-6k^2k_{0}^2)}{3(K^2)^2}\right)\,.
\end{equation}

\subsection{Summary of the calculations}

In the previous sections we have shown how the effects of a small fermion mass can be incorporated in the calculation of the power corrections of the photon polarization tensor in a hot and dense medium. From the results of the previous section, we can write
\begin{equation}\label{longitudinal_full}
   \Pi^{\text {Pow }}_{\mathrm{L}}(K)=\Pi^{\text{P0}}_{\mathrm{L}}-K^2\frac{e^2}{384 \pi^2}\frac{m^4}{T^4}\left(\text{Li}^{(1)}_{-4}(-e^{\frac{-\mu}{T}})+\text{Li}^{(1)}_{-4}(-e^{\frac{\mu}{T}})\right)\left(1+8\frac{k^2k_{0}^2}{(K^2)^2}\right)
\end{equation}
and 
\begin{equation}\label{transverse_full}
    \begin{aligned}
    \Pi^{\mathrm{Pow}}_{\mathrm{T}}&=\Pi^{\text{P0}}_{\mathrm{T}}+\frac{k^2e^2}{24 \pi^2}\frac{m^2}{T^2}\left(\text{Li}_{-2}^{(1)}(-e^{-\frac{\mu}{T}})+\text{Li}_{-2}^{(1)}(-e^{\frac{\mu}{T}})\right)\\
    &-\frac{K^2e^2}{384\pi^2}\frac{m^4}{T^4}\left(\text{Li}^{(1)}_{-4}(-e^{\frac{-\mu}{T}})+\text{Li}^{(1)}_{-4}(-e^{\frac{\mu}{T}})\right)\left[1-\frac{\left( 5k^4+k_{0}^4+2k^2k_{0}^2\right)}{2(K^2)^2}\right]\,,
    \end{aligned}
\end{equation}
where we point out again that $\Pi^{\text{P0}}_{\mathrm{L}}$ and $\Pi^{\text{P0}}_{\mathrm{T}}$ correspond to the power corrections in a system of massless fermions, evaluated previously in Refs. \cite{Carignano:2017, Carignano:2019ofj, Carignano:2021zhu,Manuel:2016wqs, Gorda:2022}. Moreover, it is important to recall that the retarded prescription $k_{0}\rightarrow k_{0}+i\eta$ is implied in these expressions. Notice that the first mass correction to $\Pi^{\text {Pow }}_{\mathrm{L}}(K)$ is proportional to $\frac{m^4}{T^4}$, since the $\frac{m^2}{T^2}$ term of $\Pi^{\text {Pow }}_{00}(K)$ vanishes. The derivatives of polylogarithm functions with respect to their order can be replaced by a combination of zeta, gamma and digamma functions, as discussed in Ref. \cite{Gorda:2023zwy}, which may facilitate numerical computations.

In the limit of vanishing chemical potential, $\mu\rightarrow 0$, we can show that the mass corrections in \eqref{longitudinal_full} and \eqref{transverse_full} reduce to (see Appendix \ref{integrals})
\begin{equation}\label{Longituginal_hot}
   \Pi^{\text {P1}}_{\mathrm{L}}(K)+\Pi^{\text {P2}}_{\mathrm{L}}(K)=-K^2e^2\left[\frac{31\zeta'(-4)}{192 \pi^2}\frac{m^4}{T^4}\right]\left(1+8\frac{k^2k_{0}^2}{(K^2)^2}\right),
\end{equation}
and
\begin{equation}\label{Transverse_hot}
     \Pi^{\text {P1}}_{\mathrm{T}}(K)+\Pi^{\text {P2}}_{\mathrm{T}}(K)=k^2e^2\frac{7\zeta'(-2)}{12\pi^2}\frac{m^2}{T^2}-K^2e^2\left[\frac{31\zeta'(-4)}{192 \pi^2}\frac{m^4}{T^4}\right]\left[1-\frac{\left( 5k^4+k_{0}^4+2k^2k_{0}^2\right)}{2(K^2)^2}\right].
\end{equation}
In the cold and dense limit, $T\rightarrow 0$, we can take a step back and reevaluate the integral $\mathcal{R}_{5}$ in the degenerate limit (see Appendix \ref{integrals}) to find that
\begin{equation}\label{Longituginal_dense}
   \Pi^{\text {P1}}_{\mathrm{L}}(K)+\Pi^{\text {P2}}_{\mathrm{L}}(K)=K^2\frac{e^2}{4\pi^2}\frac{m^4}{\mu^4}\left(1+8\frac{k^2k_{0}^2}{(K^2)^2}\right),
\end{equation}
and
\begin{equation}\label{Transverse_dense}
    \Pi^{\text {P1}}_{\mathrm{T}}(K)+\Pi^{\text {P2}}_{\mathrm{T}}(K)=-\frac{k^2e^2}{6\pi^2}\frac{m^2}{\mu^2}+K^2\frac{e^2}{4\pi^2}\frac{m^4}{\mu^4}\left[1-\frac{\left( 5k^4+k_{0}^4+2k^2k_{0}^2\right)}{2(K^2)^2}\right].
\end{equation}

Similarly to the high-temperature case, the first correction to the longitudinal component is proportional to $\frac{m^4}{\mu^4}$ in the cold and dense case.
\end{widetext}

\section{Electromagnetic susceptibilities in hot and dense systems}\label{section4}

\subsection{Electric and magnetic susceptibilities}

The electric (magnetic) susceptibility is a measure of the linear response of a system to an applied weak external electric (magnetic) field. It has been recently proposed \cite{Endrodi:2022} that, for a system in equilibrium, the electric susceptibility would be related to the photon vacuum polarization tensor through\footnote{This susceptibility differs from the one evaluated in Ref. \cite{Stetina:2017ozh} since the susceptibilities evaluated here (and in Ref. \cite{Endrodi:2022}) are built within the assumption that the system is in equilibrium and that a density distribution is induced in order to equilibrate this system. } 
\begin{equation}\label{susceptibility_original}
    \xi=\frac{1}{2 e^2} \lim _{k \rightarrow 0} \lim _{k_0 \rightarrow 0} \frac{\partial^2 \Pi_{00}^{T \neq 0}\left(k_0, k\right)}{\partial k^2} \,,
\end{equation}
and analogously one can write
\begin{equation}
     \chi=\frac{1}{2 e^2} \lim _{k \rightarrow 0} \lim _{k_0 \rightarrow 0} \frac{\partial^2 \Pi_{S}^{T \neq 0}\left(k_0, k\right)}{\partial k^2}
\end{equation}
    for the magnetic susceptibility, where $\Pi_{S}=\sum \Pi_{ii}$. So, we see that these susceptibilities are related to $\mathcal{O}(k^2)$ contributions to $\Pi_{00}(K)$ and $\Pi_{\mu \mu}(K)$, which are the quantities that we evaluated previously for light fermions at finite temperature and chemical potential. Moreover, since the background electromagnetic field is classical, only the one-loop diagram for $\Pi_{00}$ is relevant for the susceptibility. 

In this section we apply the results derived in the previous sections to the evaluation of these susceptibilities. One subtlety here is that the power corrections are usually evaluated while taking the vacuum contributions into account throughout the whole calculation, since they cancel infrared divergences in the $m=0$ terms evaluated previously \cite{Carignano:2017, Carignano:2019ofj, Carignano:2021zhu, Gorda:2022}. We, therefore, keep the vacuum terms ($T=0$) and renormalize the results before computing the susceptibilities. Namely, instead of Eq. \eqref{susceptibility_original} we  use
\begin{equation}\label{electric_susceptibility}
    \xi=\frac{1}{2 e^2} \lim _{k \rightarrow 0} \lim _{k_0 \rightarrow 0} \frac{\partial^2 \Pi_{\text{L}}\left(k_0, k\right)}{\partial k^2},
\end{equation}
and 
\begin{equation}\label{magnetic_susceptibility}
     \chi=-\frac{1}{2 e^2} \lim _{k \rightarrow 0} \lim _{k_0 \rightarrow 0} \frac{\partial^2 \Pi_{\text{T}}\left(k_0, k\right)}{\partial k^2},
\end{equation}
where we have written the expressions in terms of the longitudinal and transverse components of the photon self-energy. The additional minus sign in Eq. \eqref{magnetic_susceptibility} is due to the different sign conventions adopted in Ref. \cite{Endrodi:2022} and here (which are the same as in Ref. \cite{Gorda:2022} and are compatible with \cite{Carignano:2017}). The magnetic susceptibility in hot systems is a well-known quantity and has been evaluated using different methods, as discussed in Ref. \cite{Endrodi:2022}, and provides a useful crosscheck. 

\subsection{Susceptibilities for a hot medium}

We can now evaluate the electric and magnetic susceptibilities of a hot noninteracting medium of massive fermions. Taking the high-temperature limit of Eqs. \eqref{Transverse_massless} and \eqref{Longituginal_massless} and using the results from Eqs. \eqref{Longituginal_hot} and \eqref{Transverse_hot}, we find 

%
\begin{eqnarray}
    \label{electric_suscep_high_temp}
    \xi &=& -\frac{1}{12 \pi^2}\left(\ln \frac{T^2 \pi^2}{\bar{\Lambda}^2}-2 \gamma_E+1\right) \nonumber \\
   && -\frac{31 \zeta^{\prime}(-4)}{192 \pi^2} \frac{m^4}{T^4}+\mathcal{O}\left(\frac{m^6}{T^6}\right) \,,
\end{eqnarray}

\begin{eqnarray}\label{magnetic_suscep_high_temp}
   \chi &=& \frac{1}{12 \pi^2}\left(\ln \frac{T^2 \pi^2}{\bar{\Lambda}^2}-2 \gamma_E\right)-\frac{7\zeta'(-2)}{12\pi^2}\frac{m^2}{T^2} \nonumber \\
   && +\frac{31 \zeta^{\prime}(-4)}{192 \pi^2} \frac{m^4}{T^4}+\mathcal{O}\left(\frac{m^6}{T^6}\right).
\end{eqnarray}
%


\begin{figure}[b]
    \centering
    \includegraphics[width=8cm]{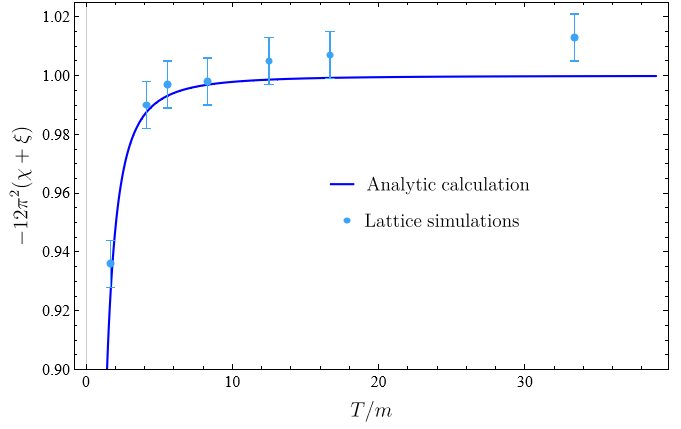}
    \caption{Sum of the electric and magnetic susceptibilities, $\xi+\chi$, obtained on the lattice \cite{Endrodi:2022} and from our analytic expressions. }
    \label{fig:Comaparrison_Suscep_lattice}
\end{figure}

Notice that, for $\Lambda \sim T$, the logarithm is very small and may even vanish. In such case, it is the mass corrections that determine the dependence of the susceptibilities with the temperature. Our approach for computing the electric susceptibility is in agreement with the results obtained in Ref. \cite{Endrodi:2023wwf}, since our computations of $\Pi_{\text{L}}$ are consistent with the massless case leading to the correct behavior of $\xi$ in that limit. Moreover the first two terms for the magnetic susceptibility are in agreement with previous calculations performed with high temperature expansions within different approaches (see Ref. \cite{Bali:2020bcn} and references therein). 


We can add the previous results, obtaining
\begin{equation}\label{sum_susceptibilities_hot}
    \xi + \chi =-\frac{1}{12 \pi^2} -\frac{7\zeta'(-2)}{12\pi^2}\frac{m^2}{T^2}+\mathcal{O}\left(\frac{m^6}{T^6}\right) \,,
\end{equation}
an observable that has been recently evaluated on the lattice\footnote{In principle, this quantity would be related to the speed of light in the medium. However, due to the imposed equilibrium condition, this is not the case here \cite{Endrodi:2022}.} \cite{Endrodi:2022}. Because of the cancellation of the logarithmic dependence, it is independent of the renormalization scale and can provide information on the mass corrections. Our calculations are similar to the findings of Ref. \cite{Endrodi:2022} and reproduce the behavior obtained on the lattice, as can be seen in Fig. \ref{fig:Comaparrison_Suscep_lattice}\footnote{There is a difference in the sign of the $\mathcal{O}(\frac{m^4}{T^4})$ term of Eq. \eqref{electric_suscep_high_temp}. We find that the $\mathcal{O}(\frac{m^4}{T^4})$ term vanishes, since the electric and magnetic contributions cancel exactly at this order. Numerical simulations still lack the sensitivity to provide information on this term.}. Our approach also has the advantage of yielding a well-defined chiral limit for each susceptibility.

\subsection{Susceptibilities in a dense medium}

The limit of cold and dense matter ($\mu\gg T$) is of general interest because of its possible implications in the physics of neutron stars. These are very dense fermionic systems in which electric and magnetic fields are known to be present, which motivates us to investigate the response of a dense noninteracting fermionic medium to an applied electric or magnetic field, using Eqs. \eqref{electric_susceptibility} and \eqref{magnetic_susceptibility}.

In the limit $T \rightarrow 0$, one can use the high chemical potential limit of Eqs. \eqref{Transverse_massless} and \eqref{Longituginal_massless}, together with our results \eqref{Longituginal_dense} and \eqref{Transverse_dense}, to show that
\begin{equation}
    \xi=-\frac{1}{12 \pi^2}\left(\ln \frac{4\mu^2 }{\bar{\Lambda}^2}+1\right)+\frac{1}{4\pi^2}\frac{m^4}{\mu^4}+\mathcal{O}\left(\frac{m^6}{\mu^6}\right) \, ,
\end{equation}
\begin{equation}
    \chi=\frac{1}{12 \pi^2}\left(\ln \frac{4\mu^2 }{\bar{\Lambda}^2}\right)+\frac{1}{6\pi^2}\frac{m^2}{\mu^2}-\frac{1}{4\pi^2}\frac{m^4}{\mu^4}+\mathcal{O}\left(\frac{m^6}{\mu^6}\right) \,.
\end{equation}

Notice that the dependence on the chemical potential is similar to the one we found for in thermal case. Moreover, as in the high-temperature limit, the dependence with the scale also drops from the sum $\xi + \chi$. Because of the absence of Fermi-distributions, the Zeta functions do not appear in the final expressions, which render the coefficients of the expansion larger. Therefore, in the dense case mass corrections bring larger contributions to the susceptibilities and their sum.

\color{black}

\section{Discussion and outlook}\label{section5}

We have discussed how to include small masses in the calculations of power corrections, calculated its first nonzero mass contributions and showed how these results have direct implications in the evaluation of electric and magnetic susceptibilities of hot and dense media in equilibrium.

As a check of our methods we computed the mass corrections to the leading-order of the  polarization tensor and confirmed that they are in agreement with the results obtained in Ref. \cite{Comadran:2021}, where the calculations are performed using on-shell effective theory and later rederived with transport theory. We have checked that their result is valid also including a nonzero chemical potential.

Notice that, in the results obtained in Refs. \cite{Carignano:2017, Gorda:2022}, in the $\mu\rightarrow 0$ limit, $\Pi^{\text {P0}}_{\mathrm{L}}$ and $\Pi^{\mathrm{P0}}_{\mathrm{T}}$ have a temperature dependence of the form $\sim \ln \frac{T}{\bar{\Lambda}}$, where $\bar{\Lambda}$ is the renormalization scale in the $\overline{\mathrm{MS}}$-scheme. Therefore, depending on the choice of the scale, this logarithm will vanish, implying that mass corrections evaluated here will determine the temperature dependence of the power corrections to the photon polarization tensor (one finds similar conclusions for the dependence on the chemical potential).

Three types of corrections to the photon polarization tensor have been investigated previously. Loop corrections (up to two loops, within the HTL approximation), power corrections and mass corrections to the one-loop polarization tensor. It has been shown \cite{Comadran:2021} that these mass, two-loop and power corrections can give contributions of the same order for masses in the soft scale. For $eT < m \ll T$ (or $e\mu < m \ll \mu)$, the mass corrections can be larger than the other kinds.

 For simplicity, let $\bar{\Lambda}=T$. Also, recall that LO HTL gives a contribution of order $e^2T^2$. Then, for $m$ and $k$ in the soft scale, i.e, $m,k\sim eT$, we see that  the power corrections are of order $\frac{k^2}{T^2}\sim e^2$ with respect to HTL, and the mass corrections computed here will give contributions of order $\frac{k^2m^4}{T^6}\sim e^6$ (longitudinal component) and $\frac{k^2m^2}{T^4}\sim e^4$ (transverse component) with respect to HTL. For $eT< m\ll T $, this contribution will be larger, but still subleading since this corrections go with negative powers of the temperature.
Therefore, for quantities in which all types of HTL corrections are taken into account, our results will be useful when one is interested in very precise calculations. Also in the case in which the coupling is not as small as in QED, since these results can be straightforwardly extended to QCD (at least the quark contribution, see discussion in Refs. \cite{Comadran:2021, Manuel:2016wqs} and the results of Ref. \cite{Ekstedt:2023oqb}). 

From the results of Ref. \cite{Gorda:2022} we may expect similar radial integrals for the mass corrections to two-loop contributions. Therefore, unless the angular integrals are shown to vanish, the LO mass dependence of the two-loop diagrams may be roughly estimated to be $\mathcal{O}(\frac{e^4m^2}{T^2})$, giving a contribution of order $\frac{e^2m^2}{T^4}\sim e^4$ with respect to HTL, which is as large as the corrections calculated in this paper. Further work is therefore needed to compute the full mass corrections at this order.

The physics of extreme systems such as heavy-ion collisions, neutrons stars or even the early universe is known to be affected by the presence of background electromagnetic fields. In contrast to magnetic effects, the computation of electric effects are still debated. The electric susceptibility proposed by \cite{Endrodi:2022,Endrodi:2023wwf} offers the possibility of performing investigations through lattice simulations. A theoretical effort to compute these same quantities is essential to advance our understanding of these electromagnetic effects. 

Since the electric and magnetic susceptibilities in hot and dense media in equilibrium are only sensitive to terms of order $\mathcal{O}(k^2)$ of the photon polarization tensor, power corrections are a crucial ingredient in their computation. Our approach offers a systematic way to compute $\mathcal{O}(\frac{m^n}{T^n})$ corrections to these quantities, while being consistent with the $m\rightarrow 0$ limit and naturally introducing a dependence with the renormalization scale. In particular, the sum of electric and magnetic susceptibilities was computed previously on the lattice \cite{Endrodi:2022}. This observable is independent of the renormalization scale, and the mass corrections evaluated here determine its temperature dependence. Results from Ref. \cite{Endrodi:2023wwf} are also in agreement with our approach since our expressions are fully consistent with the $m\rightarrow 0$ limit. Finally, we also extended the previous treatments to the cold and dense case, which can have implications in the physics of astrophysical systems such as neutron stars.

\begin{acknowledgments}
We would like to thank Cristina Manuel for helpful discussions. We also thank Gergely Endrodi and Aleksi Vuorinen for useful comments. This work was partially supported by INCT-FNA (Process No. 464898/2014-5), CAPES (Finance Code 001), CNPq, and FAPERJ.
\end{acknowledgments}

\appendix

\section{Useful integrals and identities}\label{integrals}

In this work we borrow most of the notation from Ref. \cite{Gorda:2022}. This allows us to make direct comparisons with previous results and facilitate possible applications that may include both results. Here, we introduce the basic notation and integrals that were used in our computations. We call attention to the fact that we denoted the radial integrals differently from Ref. \cite{Gorda:2022}. The integrals were computed within dimensional regularization, with the following notations 
\begin{equation}
    \begin{aligned}
    \int_P &\equiv\int_{-\infty}^{\infty} \frac{\mathrm{d} p^0}{2 \pi} \int_{\mathbf{p}}\\
    &\equiv\left(\frac{\mathrm{e}^{\gamma_{\mathrm{E}}} \bar{\Lambda}^2}{4 \pi}\right)^{\frac{3-d}{2}}\int_{-\infty}^{\infty} \frac{\mathrm{d} p^0}{2 \pi} \int \frac{\mathrm{d}^d \mathbf{p}}{(2 \pi)^d} \,,
    \end{aligned}    
\end{equation}
with the spatial integration denoted by \cite{Laine:2016book}
\begin{equation}
   \begin{aligned}
       \int_{\mathbf{p}} &=\mathcal{N}\int_0^{\infty} \mathrm{d} p p^{d-1} \int_{-1}^1 \mathrm{~d} z\left(1-z^2\right)^{\frac{d-3}{2}} .\\
       &\equiv\mathcal{N}\int_{p}\int_{z} \,,
   \end{aligned}
\end{equation}
where $z=\hat{\mathbf{k}} \cdot \hat{\mathbf{p}}$, with $\hat{\mathbf{k}}$ corresponding to an external spatial unit vector and $\hat{\mathbf{p}}$ being the spatial unit vector in the direction of the loop momentum. The normalization factor $\mathcal{N}$ was defined as
\begin{equation}
    \mathcal{N}\equiv\frac{4}{(4 \pi)^{\frac{d+1}{2}} \Gamma\left(\frac{d-1}{2}\right)}\left(\frac{\mathrm{e}^{\gamma_{\mathrm{E}}} \bar{\Lambda}^2}{4 \pi}\right)^{\frac{3-d}{2}} \,.
\end{equation}
With this notation, the radial integrals we need for our computations can be written as
%
\begin{align}\label{Dirac_Integrals}
& \int_p p^\alpha N_F^{ \pm}(p)=T^{d+\alpha} \Gamma(d+\alpha) \operatorname{Li}_{d+\alpha}\left(-\mathrm{e}^{\mp \frac{\mu}{T}}\right)\\
& \int_p p^\alpha \frac{\mathrm{d}}{\mathrm{d} p} N_F^{ \pm}(p)=-T^{d+\alpha-1} \Gamma(d+\alpha) \mathrm{Li}_{d+\alpha-1}\left(-\mathrm{e}^{\mp \frac{\mu}{T}}\right) \,.
\end{align}    

 Notice that the Fermi-Dirac integrals are usually defined for positive values of $\alpha$, and in this work we deal with negative values. However, we argue that the formulae above can still be used, since one can analytically continue both the Fermi-Dirac integrals and the polylogarithm functions to negative values of $\alpha$ by taking derivatives (see \cite{Cvijovi2009} and the discussion in the appendix of \cite{Gorda:2023zwy}).
Integrating equation \eqref{Dirac_Integrals} by parts, one can show that the following useful identities hold:
\begin{align}
    &\int_p p^{\alpha+1} \frac{\mathrm{d}}{\mathrm{d} p} N_F^{ \pm}(p)=-(d+\alpha)\int_p p^\alpha N_F^{ \pm}(p)\\
    &\int_p p^{\alpha+2} \frac{\mathrm{d^2}}{\mathrm{d} p^2} N_F^{ \pm}(p)=(d+\alpha +1)(d+\alpha)\int_p p^\alpha N_F^{ \pm}(p) \,.
\end{align}

In this work we used the following radial integrals:
\begin{equation}
\mathcal{R}_{5}\equiv\int_{p}\frac{\overline{N}_{F}}{p^5}=\frac{\text{Li}_{-2}^{(1)}(-e^{-\frac{\mu}{T}})+\text{Li}_{-2}^{(1)}(-e^{\frac{\mu}{T}})}{2 T^2}+\mathcal{O}(\epsilon)
\end{equation}
and
\begin{equation}
    \mathcal{R}_{7}\equiv\int_{p}\frac{\overline{N}_{F}}{p^7}=\frac{\text{Li}^{(1)}_{-4}(-e^{\frac{-\mu}{T}})+\text{Li}^{(1)}_{-4}(-e^{\frac{\mu}{T}})}{24 T^4}+\mathcal{O}(\epsilon)\,.
\end{equation}
In the $\mu\rightarrow 0$ limit, one can show\footnote{This can be shown using identities for the polylogarithm. See, for example, the identities in the appendix of \cite{Gorda:2023zwy}.} that these integrals simplify to
\begin{equation}
    \mathcal{R}_{5}=\frac{14 \zeta'(-2)}{T^2} + \mathcal{O}(\epsilon), \quad  \mathcal{R}_{7}=\frac{31 \zeta'(-4)}{12T^4}+\mathcal{O}(\epsilon).
\end{equation}
In the $T \rightarrow 0$ limit, the Fermi-Dirac integrals become step functions for particles, while for antiparticles they vanish. Thus:
\begin{equation}
    \mathcal{R}_{5}=-2\mu^{-2} +\mathcal{O}(\epsilon), \quad \mathcal{R}_{7}=-4\mu^{-4}+\mathcal{O}(\epsilon).
\end{equation}
%


Most of the angular integrals we encountered in our calculations of the form \cite{Gorda:2022, Laine:2016book}
\begin{equation}\label{angular_integrals}
   \begin{aligned}
       \mathcal{A}_\alpha &\equiv \int_z \frac{1}{(v \cdot K)^\alpha}\\
&=\frac{\Gamma\left(\frac{1}{2}\right) \Gamma\left[\frac{1}{2}(d-1)\right]}{\Gamma\left(\frac{d}{2}\right)}\left(-k^0\right)^{-\alpha}{ }_2 F_1\left(\frac{\alpha}{2}, \frac{1+\alpha}{2} ; \frac{d}{2} ; \frac{k^2}{k_0^2}\right).
   \end{aligned}
\end{equation}
In our computations we had expressions with factors $v'\cdot K=-k^0+\frac{p}{E_{p}}k z$ in the denominator. After the expansions in small mass we found factors $v\cdot K=-k^0+k z$ for the external four-momentum $K$. These are the ones that appear in \eqref{angular_integrals}.

We also encountered the integrals
\begin{equation}\label{angular_integrals_A_2_4}
    \begin{aligned}
        &\mathcal{A}^{(2)}_{4}= \int_z \frac{(kz)^2}{(v \cdot K)^4}=-\frac{2k^2(3k^2+k_{0}^2)}{3(K^2)^3}+\mathcal{O}(\epsilon)\, ,\\
        &\mathcal{A}^{(2)}_{5}= \int_z \frac{(kz)^2}{(v \cdot K)^5}=-\frac{2k^2k_{0}(5k^2+k_{0}^2)}{3(K^2)^4}\, ,\\
        &\mathcal{A}^{(2)}_{6}= \int_z \frac{(kz)^2}{(v \cdot K)^6}=\frac{-2k^2}{15}\frac{(5k^4+38k^2k_{0}^2+5k_{0}^4)}{(K^2)^5}\,,
    \end{aligned}
\end{equation}
and the identity
\begin{equation}
    \int_z \frac{kz}{(v \cdot K)^\alpha}=\left(\mathcal{A}_{\alpha-1}+k^0 \mathcal{A}_\alpha\right)\,,
\end{equation}
was frequently useful. 


\bibliographystyle{apsrev4-1}
\bibliography{references.bib}

\end{document}